\documentclass[usegraphicx,usenatbib,useAMS]{mn2e}
\newcommand{\beq}{\begin{equation}}
\newcommand{\eeq}{\end{equation}}
\newcommand{\be}{\begin{equation}}
\newcommand{\ee}{\end{equation}}

\def\gap{\;\rlap{\lower 2.5pt
 \hbox{$\sim$}}\raise 1.5pt\hbox{$>$}\;}
\def\lap{\;\rlap{\lower 2.5pt
   \hbox{$\sim$}}\raise 1.5pt\hbox{$<$}\;}

\begin{document}

\title[The abundance of dark galaxies]{The abundance of dark galaxies} 
\author[Verde, Oh \& Jimenez] 
{Licia Verde$^{1,2}$, S. Peng Oh$^3$ and Raul Jimenez$^1$ \\ 
$^1$Department of Physics and Astronomy, Rutgers University, 
Piscataway, NJ 08854--8019 USA.\\
$^2$Princeton University Observatory, Princeton, NJ 08544, USA.\\ 
$^3$California Institute of Technology, Mail Code
130-33, Pasadena, CA 91125, USA.}

\maketitle
 
\begin{abstract}
  
We show that gas in a large fraction of low mass dark matter halos  may form
Toomre stable disks, if angular momentum is conserved when the gas
contracts. Such halos would be stable to star formation and therefore  remain
dark. This may potentially explain the discrepancy between the  predicted and
observed number of dwarf satellites in the Local Group, as well as  the
deviation between the predicted and the observed faint end slope of  the
luminosity function. The above mechanism does not require a strong  variation
of the baryon fraction with the virial mass of the dark halo. We show  that
model fits to rotation curves are also consistent with this  hypothesis: none
of the observed galaxies lie in the region of parameter space  forbidden by
the Toomre stability criterion.

\end{abstract}


\section{Introduction}
There is now growing evidence that the observed number of dwarf galaxies
surrounding the Milky Way or M31 does not agree with the predictions of
high--resolution N-body simulations for the currently favored Cold Dark Matter
cosmological model (flat, low matter-density Universe -- LCDM)
\citep{KKVP99,MGGLQST99}. Given the successes of the current LCDM paradigm in
reproducing many of the large-scale observables (e.g.,
\citet{Jaffe+01,Peacock+01,Verde+2dF01,Lahav+2dF01}), it seems natural to try
to make changes to the existing paradigm without modifying its large-scale
properties.  There are two routes in this direction: one is to prevent the
formation of small dark matter halos, the second is to hide them by
suppressing star formation.  Some modifications belonging to the first group
consist of changing the nature of the dark matter itself to have different
physical properties: self--interacting \citep{SS00}, warm \citep*{CDW96,BOT01}
or other variants (e.g., \citet*{Goodman00,HBG00,Cen01}).  It has also been
proposed that the solution to the problem may lie within the nature of the
initial conditions and that some modification to the shape of the inflaton
potential may suppress small scale structure \citep{KL00}, although this
solution faces some problems in reproducing the observed amount of structure
in the Lyman-$\alpha$ forest. On the other hand, the solution of suppressing
star formation via supernova feedback was suggested a long time ago (e.g.,
\citet{WR78}).  This process expels the gas from the halo and therefore
suppresses star formation. The accretion and cooling of gas can also be
suppressed in the presence of a strong photo--ionizing background. This route
has also been known for a number of years now \citep*{DZN67} and has been
investigated recently in the context of semi--analytic models of galaxy
formation \citep{BKW01,CGO01,S01,BLBCF01}.

A closely related problem is the fact that the observed faint end of the
luminosity function has a shallower slope \citep{Blantonetal01} than predicted
from high-resolution LCDM simulations, if one simply assumes a constant
mass-to-light ratio (e.g., \citet{Jenkins+01}). Stellar feedback is again
often used in semi--analytic models to suppress star formation in halos with
shallow potential wells (e.g., \citet{Coleetal00}).

All of the above models seek to systematically increase the mass-to-light
ratio in small halos, by depleting the cold gas fraction available for star
formation. In this paper we show that such drastic expulsion of gas is,
perhaps, unnecessary: provided that the gas conserves angular momentum during
collapse, a large fraction of low mass halos will be Toomre stable
\citep{Toomre_64,Kennicutt_89}, i.e., $Q=\Sigma_c/\Sigma > 1 $ (where $\Sigma$
is the disk surface mass density and $\Sigma_c$ is the critical surface
density to trigger gravitational instability and thus start formation). If the
baryon fraction ($f_d$) in disks is $f_d \sim 0.5 \Omega_{b}/\Omega_{m}$, all
disks in halos with masses $M < 10^{9} \, M_{\odot}$ will be Toomre stable and
fail to form stars. The ``missing'' dwarf galaxies predicted to be around the
Milky Way and M31 may fall in this category.  We also show that, if a small
fraction of baryons initially present in the dark matter halo settle in a
disk, as some observational evidence suggests
\citep*{JVO02,GuzikSeljak02,vanBBS01,Burkert00,FHP98}, even more massive halos
have some probability of remaining dark, provided they have high spin.  This
could potentially reconcile the observed luminosity function with the mass
function predicted in high--resolution LCDM N-body simulations. We show that
the predictions of this model are consistent with our previous analysis of the
rotation curves of galaxies (\citet*{JVO02}, hereafter JVO02): none of the
galaxies lie in the region of parameter space forbidden by the Toomre
instability criterion.

Throughout this paper we assume a LCDM cosmology given
by:$(\Omega_{m},\Omega_{\Lambda},\Omega_{b},h,\sigma_{8}
  h^{-1})=(0.3,0.7,0.039,0.7,1.0)$.

\section{The mass distribution of dark galaxies}

We model spiral galaxies as exponential disks embedded in a cold dark  matter
halo.  Following \citet{DSS97,JHHP97,MMW98} we assume a NFW  \citep*{NFW97}
profile $\rho \propto [(r/r_c)(1+r/r_c)^2]^{-1}$ where $r_c$ is the  break
radius and the surface mass density of the disk is given by  $\Sigma=\Sigma_0
\exp[-r/Rd]$, where $R_d$ is the scale length of the disk.

For a baryonic disk to be locally gravitationally unstable, despite the
stabilizing effects of tidal shears and pressure forces, we require the Toomre
parameter $Q <1$, where:

\begin{equation}
Q= \frac{c_{\rm s} \kappa}{\pi G \Sigma},
\label{toomre}
\end{equation}

$\Sigma$ is the disk surface mass density, $\kappa = 1.41 (V/r) (1
+ {\rm d \, ln} V/{\rm d \, ln} r)^{1/2}$ is the epicyclic
frequency and $c_s$ is the gas sound speed. 

Regions in local disk galaxies where $Q>1$ are observationally  associated
with very little star formation, indicating that the Toomre criterion is
obeyed remarkably well, if a gas sound speed of $c_{s} \sim 6\, {\rm km \,
s^{-1}}$ is assumed \citep{Kennicutt_89}. \citet{FWGH96} noted that for one
galaxy  (NGC 6946) the agreement with the Toomre criterion was not perfect:
there  was no truncation of star formation for $Q > 1.6$ if the observed
vertical  velocity dispersion was adopted. However, agreement with the Toomre
criterion  could be obtained if the velocity dispersion was assumed to be
constant with  radius.  \cite{WongBlitz01} have measured the value of $Q$ in
the inner part of rotation curves for a sample of 8 galaxies. For half of the
sample the predicted value from Toomre theory for the abundance of gas is in
excellent agreement with observations. In the other cases the
agreement is not as remarkable, but still there is no evidence for large
deviations from the Toomre stability predictions.

\citet{JHHP97} developed a disk model to study the dependence of $Q$ on  mass,
radius and spin parameter of the dark matter halo and found that for a
certain range of the above parameters, some disks would be dark. An important
ingredient in their study was the assumption of a constant baryon  fraction,
namely they adopted the nucleosynthesis value from \citet{WSKSO91}.  Using
this constant disk fraction for all galaxies, they concluded that for an
Einstein--de Sitter Universe the minimum dark matter halo mass for a  visible
disk galaxy is about $10^9$ M$_{\odot}$. On the other hand, for a low  matter
density universe (e.g., LCDM), this minimum mass drops substantially,  by
about two orders of magnitude, and thus the fraction of dark halos also falls
drastically. The reason for this is that for a low density Universe  with
baryon density fixed by nucleosynthesis, the baryons constitute a  higher
fraction of the halo mass.

It seems unlikely that all baryons will be able to cool and form stars.
Recent studies from galaxy--galaxy lensing from the Sloan Digital Sky  Survey
(SDSS) show that for $L^{*}$ galaxies the baryon fraction is about half  of
the nucleosynthesis value \citep{GuzikSeljak02}. On the other hand,
theoretical considerations suggest that the baryon fraction for galaxies,
either less massive or more massive than  $L_{*}$, is likely to be much
smaller due to feedback and inefficient cooling  respectively (e.g.
\citet{vdB01,BFLBC01}). This implies that a large fraction of  disks should be
Toomre stable. Below we quantify how this affects the  predicted number
density of observable galaxies.

\begin{figure}
\includegraphics[width=9.5cm]{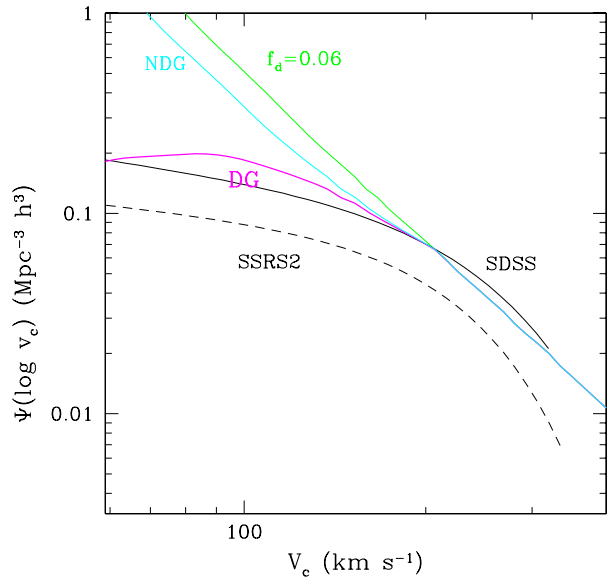}
\caption{The present day velocity function. Lines labeled by Sloan and 
SRSS2 are obtained from the SDSS and the SRSS2 surveys respectively.  The 
solid line labeled $f_d=0.06$ is the theoretical velocity function derived 
from the \citet{ST99} mass function using a constant disk baryon fraction
  $f_d=0.06$.  The line labeled NDG (no dark galaxies) is obtained 
assuming our ansatsz for the dependence of $f_d$ on mass (see figure 2). 
Finally the thick solid line labeled by DG (dark galaxies) is obtained 
considering that only disks that are Toomre unstable will be able to form
stars and  thus be visible (see text for more details).}
\label{fig:lfcomp}
\end{figure}

\subsection{Luminosity function}

CDM models typically predict many more galaxies at the faint end of the
luminosity function than are actually observed. This can most easily be 
seen by comparing the galaxy circular velocity function with theoretical
predictions.  The galaxy velocity function is obtained by combining the
observed luminosity function with empirically determined 
luminosity--velocity relations (Tully-Fisher relation for spirals, and
Faber-Jackson for ellipticals). \citet{Gonzalez+01} find that treating the
entire galaxy population as spirals does not significantly alter the derived
velocity function. In fig.~\ref{fig:lfcomp}, we plot a representative velocity 
function from their paper, derived from the SSRS2 \citep{Marzke+98} B band 
survey and the \citet{Yasuda+97} Tully-Fisher relation. 
We also construct a new  velocity
function from the Sloan Digital Sky Survey (SDSS), which has published
luminosity functions in 5 bands \citep{Blantonetal01}. We choose to use 
the R band, as the R band Tully-Fisher relation is tighter and has less 
scatter than the Tully-Fisher relation at shorter
wavelengths. \citet{Blantonetal01} perform the color transformations from SDSS
magnitudes to LCRS \citep{Shectman+96} R-band magnitudes (see their Table 3);
we perform  the color transformations and corrections for internal extinction
as in \citet{Gonzalez+01} to the $R_{\rm courteau}$ magnitudes used by
\citet{Courteau97}. We then use the \citet{Courteau97} $R_{\rm 
courteau}$ band Tully-Fisher relation to obtain the velocity function. Note
that the  Sloan velocity function has a significantly higher normalization
than the  SSRS2 velocity function. This is due to the use of Petrosian
magnitudes for  SDSS, which allows detection of galaxies of significantly
lower surface  brightness.
The luminosity density of galaxies in the SDSS is therefore 
significantly higher than other redshift surveys. The SDSS luminosity function
is  very similar to the luminosity functions derived by other surveys such as 
the Las Campanas Redshift Survey (LCRS) and Two Degree Field Galaxy Redshift 
Survey (2dFGRS) if they re-analyze their data using the shallower isophotal 
limits for galaxy magnitude employed by these surveys; see 
\citet{Blantonetal01} for discussion.

The theoretical velocity function is given by:
\begin{equation}
\Psi(V_c)=\frac{dN}{d \log V_c}= \frac{dN}{dM} \frac{dM}{d \log(V_c)} 
f_{\rm bright},
\end{equation}
where $V_c$ is the circular velocity in the flat part of the rotation curve
and $f_{\rm bright}$ is the  fraction of
halos which form stars. For the mass function $dN/dM$, we use the 
\citet{ST99} mass function, which gives a good fit to the mass function seen
in the  Hubble Volume simulations (see \citet{Jenkins+01}). For evaluating the 
derivative $\frac{dM}{d \log(V_c)}$, we use the fitting formula of
\citet*{MMW98},  which gives the circular velocity $V_c$ as a function of the
virial mass of  the dark
halo ($M_{200}$), the concentration parameter ($c$), the spin of the 
dark halo ($\lambda$) and $f_d$.  
As in \citet*{NFW97}, for a given $M_{200}$ we
evaluate the concentration parameter from the typical collapse 
redshift. Below we discuss our choice of $f_{d},\lambda$ further and evaluate 
$f_{\rm  bright}$ using Toomre instability arguments.

\begin{figure}
\includegraphics[width=8.5cm]{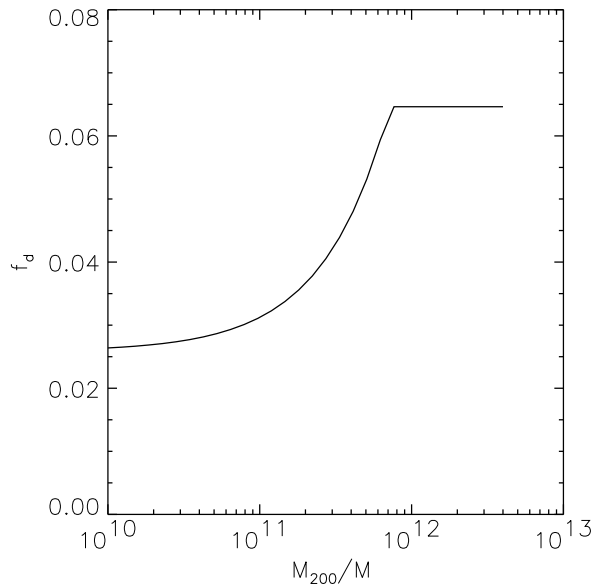}
\caption{Our ansatz for the mass dependence of the disc baryon fraction
  $f_d$. This models a factor of two decline in the value of $f_{d}$ 
from high mass to low mass halos, which fits the median value of $f_{d}$ at 
high  masses \citep*{MMW98} and low masses \citep{vanBS01}.}
\label{fig:fdm200}
\end{figure}

Let us start by assuming $f_{\rm bright}=1$, a constant disk mass 
fraction $f_{d}=0.06$ (i.e., about half the nucleosynthesis value 
\citep{FHP98}), and the disk spin parameter to be the mean value found in N-body simulations
$\lambda_{\rm mean}= 0.042$ \citep{BDKKKPP01}. In figure 
~\ref{fig:lfcomp} we plot the theoretical velocity function (upper solid line
labeled by $f_d=0.06$). At low velocities, the predicted number density of 
galaxies is much larger than observed, which is the discrepancy we seek to
resolve. 

This cannot be resolved by making $f_{d}$ a function of $M_{200}$,  unless the
$f_d(M_{200})$ dependence is unrealistically drastic.   To illustrate  this we
assume a different functional form for $f_d(M_{200})$, as shown in
fig.~\ref{fig:fdm200}.  Our {\it ansatz} for the $f_{d}(M_{200})$ relation
models a fairly mild  factor of two decline in the value of $f_{d}$ from high
mass to low mass  halos, which fits the median value of $f_{d}$ at high masses
\citep*{MMW98} and low  masses \citep{vanBS01}. If this was the only process
at play it would produce  only a mild change in the theoretical velocity
function, as shown by the thin  solid line labeled by NDG (no dark galaxies)
in figure ~\ref{fig:lfcomp}.

To include the effects of the Toomre instability criterion we proceed as
follows. We define a critical spin parameter $\lambda_{\rm crit}$ such that
the disc is marginally stable: $Q[f_{d}(M_{200}),\lambda_{\rm crit}, M_{200},
c(M_{200})]=1.5$, where $Q$ is evaluated at one disk scale-length, and we
assume a gas sound speed $c_{s}=6\, {\rm km \,s^{-1}}$ (the gas will in
general be photoionized by the meta-galactic ionizing background to $T \sim
10^{4}$K).  The fraction of bright galaxies is given by $f_{\rm
  bright}=\int^{\lambda_{\rm crit}}_{0} p(\lambda) d\lambda$, where
$p(\lambda)$ is the probability distribution of $\lambda$ derived from
numerical simulations \citep{BDKKKPP01}. Note that since halos with $\lambda >
\lambda_{\rm crit}$ are assumed not to host visible galaxies, the probability
distribution of $\lambda$ in observed galaxies is skewed. We evaluate the
circular velocity assuming a median value $\lambda_{\rm median}$, defined by
the implicit equation $\int^{\lambda_{\rm median}}_{0} p(\lambda)
d\lambda/f_{\rm bright}=0.5$. The velocity function we obtain from computing
$f_{\rm bright}$ in this manner is shown in figure ~\ref{fig:lfcomp} by the
thick solid line labeled by DG (dark galaxies).  Thus, once we include the
suppression of high-spin galaxies by explicitly computing $f_{\rm bright}$,
the theoretical predictions match the observations well.

The suppression of accretion after reionization
\citep{BKW01,CGO01,S01,BLBCF01}, is an attractive mechanism for 
regulating $f_{d}$ since (unlike for instance, supernova explosions) it
affects dark  galaxies as well. 
We model this effect by using the fitting formula \citep{Gnedin00,S01}:
\begin{equation}
f_{d}^{\rm photo}(M_{200},z_{f})=\frac{\Omega_{b}}{\Omega_{m}} 
\frac{1}{[1 +0.26
\frac{M_{\rm C}}{M_{200}}]^{3}} 
\label{eq:squelching}
\end{equation}
where we approximate $M_{C}$ as the mass corresponding to a halo with 
virial velocity $V_{c}=50 \, {\rm km \, s^{-1}}$ \citep{S01} at the redshift 
of formation $z_{\rm f}$. For a given halo mass $M_{200}$, extended
Press-Schechter theory \citep{LC94} gives the probability distribution
of collapse redshifts $p(z_{f})$; we can then use equation
(\ref{eq:squelching}) to obtain the probability distribution 
$p(f_{d})$. Note that $f_{d}=\Omega_{b}/\Omega_{m}$ for $z >
z_{\rm reion}$, where $z_{\rm reion}$ is the redshift of reionization;
we assume $z_{\rm reion} \sim 7$ (the results depend only very weakly
on $z_{\rm reion}$). 

In the photoionization model, the fraction of dark galaxies can be 
computed via equation \ref{eqn:fdark}, with the probability distribution 
$p(f_{d})$ given above. Significant suppression of the number density of low 
circular velocity galaxies only takes place off the scales plotted in fig
\ref{fig:lfcomp}. Photoionization only successfully suppresses gas 
accretion in very small halos; it succeeds in alleviating the dwarf satellite 
problem, but leaves unaffected halos of $V_c > 75$ Km s$^{-1}$. By itself it 
therefore cannot produce the low values of $f_{d}$ required to reconcile the 
theoretical and observed velocity function.

We do not attempt to model the $f_{d}(M_{200})$ relation in more detail 
(which is the task of hydrodynamic galaxy formation modelling), but merely 
note that it is consistent with observations and theoretical prejudice that gas
expulsion and suppression of accretion are more important for shallow
potential wells. Possible mechanisms include suppression of accretion 
due to entropy injection in the intergalactic medium by supernovae and/or AGN 
jets, or ram pressure stripping as a smaller halo falls into a larger halo. 
We note merely that by invoking the Toomre instability criterion, we do not 
require the very strong variation of $f_{d}$ with $M_{200}$ required in most
semi-analytic models of galaxy formation. In particular, we do not 
require a very sharp drop in $f_{d}$ at the faint end.

\begin{figure}
\includegraphics[width=8.5cm]{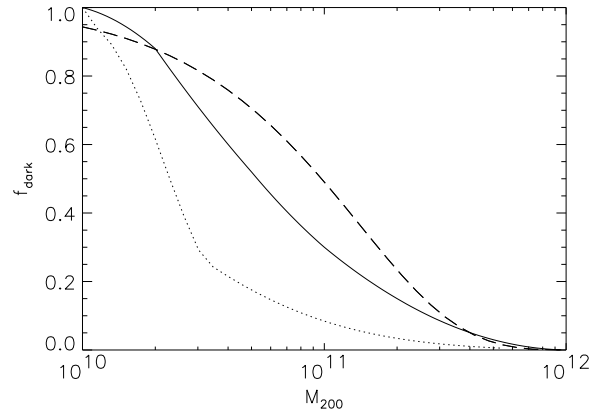}
\caption{Fraction of dark galaxies $f_{\rm dark}$ as a function of halo
mass. The dashed line is $f_{\rm dark}=1-f_{\rm bright}$, where $f_{\rm
bright}$ has been computed  in \S 2, by considering the Toomre instability
criterion and a mild mass  dependence of the disk baryon fraction $f_d$ shown
in fig \ref{fig:fdm200}. The  dotted line shows the resulting $f_{\rm dark}$
if suppression of accretion  due to reionization determines $f_{d}$. The solid
line is the dark fraction  derived using the distribution of $f_d$ empirically
obtained by fitting the  rotation curves.}
\label{fig:fdark}
\end{figure}

\subsection{The fraction of dark galaxies}

In figure~\ref{fig:fdark} we plot the fraction of dark galaxies $f_{\rm 
dark}$ as a function of halo mass.

The dashed line is $f_{\rm dark}=1-f_{\rm bright}$, where $f_{\rm 
bright}$ has been computed in the previous section.  The dotted line shows the 
resulting $f_{\rm dark}$ if photoionization was the only process at play. As 
already noted this mechanism only suppresses accretion in halos of low circular
velocity and cannot reconcile the observed and theoretical velocity 
function.
Assuming $f_d$ as a deterministic function of $M_{200}$ might be an
oversimplification: in reality $f_{d}$ is likely to be highly 
stochastic (e.g., \citet{vdB01,BFLBC01}). Also, analysis of rotation curves 
(JVO02) confirms that there is a significant scatter in the relation (see \S 
3). We thus compute also the expected fraction of dark galaxies by neglecting 
any possible mass dependence and using the distribution of $f_d$ ($p(f_d)$)
empirically obtained by fitting the rotation curves for the NFW profile 
(see JVO02 and \S 3):
\begin{equation}
f_{\rm dark}=\int_{\lambda_{crit}}^{\infty}d\lambda \int_{0}^{f_{d\; 
{\rm  crit}}}df_d \, p(f_d) \, p(\lambda)
\label{eqn:fdark}
\end{equation}
where $\lambda_{\rm crit}$ and $f_{d\; {\rm crit}}$ are defined by 
$Q(M_{200}, f_{d\; {\rm crit}},\lambda_{\rm crit},c)=1$.  The result is shown
in  figure
\ref{fig:fdark} as a solid line.  The two approaches (the one described 
in \S 2.1 or using the empirically obtained $f_d$ distribution) yield very 
similar values for $f_{\rm dark}$, indication of the fact that the main  mechanism to create dark galaxies is the Toomre criterion.

\begin{figure}
\includegraphics[width=8.5cm]{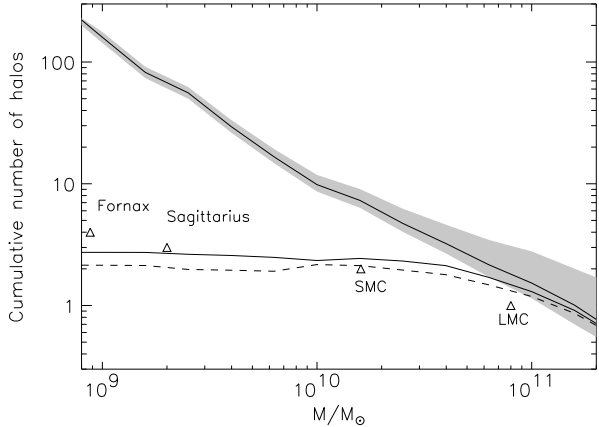}
\caption{Cumulative mass function of dwarf satellites of the Milky Way. Small
triangles are the observations, the solid line with the  gray area is the
predictions from N-body simulations with an estimate of the
uncertainty. Finally the solid and dashed dashed lines show the  predictions
once the the probability that some halos will remain dark (due to  their spin
parameter being above the critical value) has be taken into account.  Solid
and dashed lines have been computed using the $f_{\rm dark}$ fraction  of
fig. 4. }
\label{fig:cumulativedark}
\end{figure}

\subsection{Dwarf satellites}
From fig.~\ref{fig:fdark} it is clear that star formation within dark  matter
halos with masses below $10^{10}$ M$_{\odot}$ will be almost completely
suppressed (not visible because they will be Toomre stable). Thus this
provides a mechanism to hide these low mass halos.  To make a more
quantitative estimate of the viability of this effect, in figure
\ref{fig:cumulativedark} we compare the observed cumulative mass  function of
dwarf satellite with predictions from N-body simulations.  The solid  line is
from \citet{MGGLQST99}, where the velocity has been converted in mass.  The
gray area is an estimate of the error from fig. 2 of \citet{MGGLQST99}.   The
solid and dashed lines are the predicted cumulative number of  satellites once
the probability that some galaxies will remain dark has be taken into
account.  In particular we have used the predictions of the solid and dashed
lines of figure \ref{fig:fdark} respectively.  This prediction agrees nicely
with observations (triangles).  We have not shown results for masses below
$10^9$ M$_{\odot}$ because the uncertainties in the procedure to obtain the
corrected line become too important; only a very small fraction ($< 1$\%) of
dark  matter halos will be visible thus making this fraction very dependent on
the  exact value of the predicted number of dark galaxies.

\section{Rotation Curves}
In this section we present some indication of the  existence of dark  galaxies
from analysis of a large sample of rotation curves.

We use the same set of rotation curves as in JVO02 where we determined  the
best fit disk parameters within the context of both a disk within a NFW
profile halo and within a pseudo-isothermal profile halo (see JVO02 for
details).

The two sets of observed rotation curves are:

\begin{itemize}
\item Courteau
catalog \citep{Courteau97} which consists of optical ($H_{\alpha}$)
long--slit rotation curves for 300 Sb--Sc UGC galaxies.
\item 64 spiral
galaxies (Sa--Sd) observed in $H_{\alpha}$ by
\citet{PW00}
\end{itemize}

In total we use 364 optical rotation curves (to avoid any problems with the
smearing of the beam in radio observations  \citep{vanBS01}) corrected for
inclination and with error bars. For  these galaxies the scale length of the
exponential disk ($R_d$) is also  measured.  The two models have three free
parameters: $M_{200}$, $r_c$, and $f_d$.  In principle there is another disk
parameter: the modified spin  parameter $\lambda'$. If $\lambda$ is the spin
parameter of the dark halo, then $\lambda'=j_d/f_d\lambda$ where $j_d\equiv
J_d/J$ and $J_{d},J$ are the  total angular momentum of the disk and halo
respectively.  $\lambda'$ can be  readily obtained using the measured value of
$R_d$ and the best fit parameters  (see JVO02). If angular momentum is
conserved, then $\lambda=\lambda'$, i.e.  $j_d/f_d=1$.

In JVO02 we used the sample of 364 galaxies for which $R_d$ is  measured, we
found the best fitting parameters ($M_{200}$, $f_d$, and $c$) for the  above
two models and recovered $\lambda'$.  Within both models for the  majority of
the galaxies, a meaningful set of fitting parameters was recovered. For  the
remaining galaxies we obtained values for $f_d$ that where either  implausibly
low ($f_d \simeq 0.001$) or may be too high ($f_d \gap 0.2$).

\subsection{Fit to the disk models}

As described in detail in JVO02, for each galaxy rotation curve and for  both
models, the best fitting parameters were obtained using a standard  $\chi^2$
minimization and exploring the whole likelihood surface.  The sample of
galaxies studied is very non-uniform: some curves have much smaller
error-bars than others, some curves do not extend to large enough radii to
show  the flattening of the rotation curve, while others show strong evidence
of  spiral arms and bars in the rotation pattern, moreover we have not
attempted  to model any bulge nor bar component.  Thus it is is important to
keep in mind  that galaxies might be more complicated than our model for
galaxy rotation  curves.  However, for most of these galaxies, the
high-quality of the rotation  curves measurements allow degeneracies among the
disk parameters to be lifted (JVO02). Here, we are not interested in a
detailed modeling of  individual galaxy dynamics but rather the general
statistical trends of the  recovered disk parameters from the whole sample.

\begin{figure*}
\includegraphics[width=17cm,height=15cm]{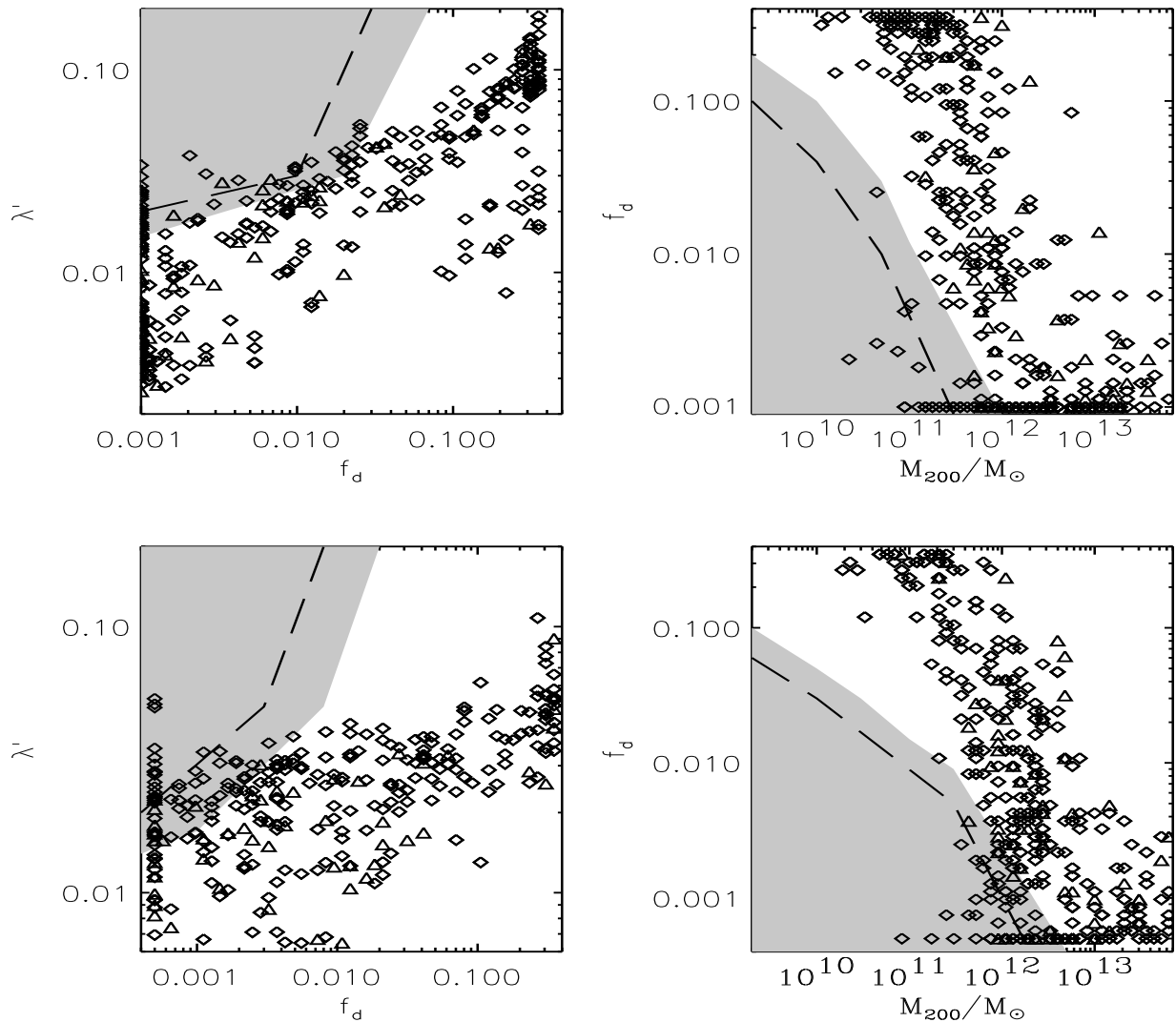}
\caption{Distribution of the best fitting disk parameters in the
$f_d$-$\lambda'$ plane and in the $f_d-M_{200}$ plane for the NFW (top) and
pseudoisothermal profile (bottom). We argue that the  Toomre instability
criterion correctly predicts the zone of avoidance (gray   shaded areas) for
combinations of disk parameters that involve low $M_{200}$   and low $f_d$.
Note that there seems to be another small void region in  the  bottom right
corner of the $f_d$-$\lambda'$ plot. This might arise  because  for these
parameter values discs might be too concentrated and form a spheroidal
instead of a disc.}
\label{fig:darkbestfits}
\end{figure*}

Fig.~\ref{fig:darkbestfits} shows the distribution of the best fitting  disk
parameters in the $f_d$-$\lambda'$ plane and $M_{200}$-$f_d$ plane, for  the
NFW profile (top) and pseudoisothermal profile (bottom); diamonds  correspond
to galaxies from the \citet{Courteau97} sample and triangles are from
\citet{PW00}.  In JVO02 we showed that low surface brightness galaxies  (LSBs)
follow the same trend as the rest of the sample, so any observational  bias
against LSBs does not seem to affect this result: the only  difference is
that LSBs have systematically higher values of $\lambda'$ for a fixed  value
of $f_d$.  In both models there is a correlation between $\lambda'$ and
$f_d$, albeit with a scatter, and a weak anti-correlation between $M_{200}$
and $f_d$.

For $f_d < 0.02$, $\lambda'$ values are between 0.01 and 0.03, while  for $f_d
> 0.02$, $\lambda'$ takes all values between 0.02 and 0.4.

The origin of this correlation is controversial: \citet{vdB01} argues  that
when feedback by star formation is included in models of galaxy  formation,
such a correlation between $\lambda'$ and $f_d$ arises naturally for  low mass
galaxies, where feedback is most efficient; \citet{Burkert00} showed  that it
can be an effect of disk growth happening inside-out; or disks might
preferentially lose baryons with high angular momentum.  On the other  hand,
\citet*{BvdBS02} argue that the correlation may be entirely due to  intrinsic
degeneracies due to the fact that from rotation curves analysis the  total
dark halo mass is not well constrained.  In JVO02 (figure~14) we find that
this is the case for rotation curves that are not measured at large enough
radii, but, for most of the galaxies in the sample (70\%), the high-quality
rotation curves allow model degeneracies to be lifted. This indicates that the
random error in recovered parameters is small. On the other hand, the
systematic errors (especially due to the uncertainty in the dark matter
profile)  may be larger, possibly causing the correlations.

We argue here that the most significant feature of the left panels of  fig. 1
{\em is not the correlation itself}, but the fact that spirals avoid  the
upper left portion of the $\lambda'$--$f_d$ plane and that this is in
agreement with the Toomre instability criterion.  In passing we also note that
disk instability might make galaxies with disk parameters in the bottom left
corner of the plot to drop out of the sample: for these parameter values discs
might be too concentrated and form a spheroidal instead of a disc or galaxies
might become bulge dominated.

The right panels show an interesting, although weak, correlation   between
$M_{200}$ and $f_d$: low mass spirals ($M_{200} < 10^{11} M_{\odot}$)   tend
to have $f_d > 0.1$ since no galaxies are found in the lower left portion   of
the plot.  There can be several different effects that module the   dependence
of $f_d$ on the halo mass.  Linear theory evolution from inflation   predicts
a weak anti correlation between spin parameter and peak height   \citep{HP88},
and hence mass of the halo. The expected anti correlation is, however, very
weak.  On the other hand, cooling is less efficient in massive halos ($M_{200}
\gap 10^{12} M_{\odot}$), thus $f_d$ is expected to {\it decrease} strongly
with increasing $M_{200}$ (e.g., \citet{vdB01}) at the high mass end.
However, in small mass halos, reheating of cooled gas through feedback and
photoionization processes are much more efficient than in massive systems,
thus   implying that $f_d$ should {\it increase} with $M_{200}$ (e.g.,
\citet{BFLBC01}) at   the low mass end. The correlation we find from rotation
curves analysis is in   rough agreement with the high-mass end trend (see
fig.~5 of \citet{vdB01}).   For his no feedback model, the baryon fraction
decreases for increasing masses.   For the feedback model this is only true
for $M_{200}$ bigger than $\sim 5   \times 10^{11}$ M$_{\odot}$. For lower
masses this trend reverses and we do   not find evidence for it from our
analysis of rotation curves.  This may be due   to the fact that for these low
masses, only those galaxies with highest $f_d$   are visible.

Here we show that Toomre instability criterion predictions are in  agreement
with the zone of avoidance for combinations of disk parameters that  involve
low $M_{200}$ and low $f_d$ or low $\lambda$. 

\subsection{Stability of discs}

The void areas found in fig.~\ref{fig:darkbestfits} can be explained by
resorting to the Toomre instability criterion: for $Q > 1 $ the gas  disc is
stable and thus star formation is strongly suppressed.  Expressions for  $Q$
and $\Sigma$ for the NFW profile have been presented in \citet{JHHP97},  here
we derive the corresponding expressions for the pseudoisothermal  profile.

Recall that in the pseudoisothermal profile the density $\rho$ as a function
of the radius $r$ is $ \rho(r)=\rho_0 [1+ (r/rc)^2]^{-1} $, where $\rho_0$
denotes the central finite density.  The disk surface density as a function of
the radius is $\Sigma (r) =\Sigma_0 \exp(-r/R_d)$, where $\Sigma_0=f_d
M_{200}/(2 \pi R_d^2)$.  Appendix I of JVO02 gives a relation between $\rho_0$
and the concentration parameter $c$ and a fitting formula for $R_d$ as a
function of the disk parameters: $M_{200}$, $\rho_0$, $r_c$ and $\lambda'$.

The shaded areas in figure \ref{fig:darkbestfits} represent the void  regions
predicted by the Toomre criterion assuming $c_s= 6$ km s$^{-1}$ for all
values of masses and radii. The dashed line shows how the avoidance region
would change if $c_s=2$ km s$^{-1}$. Such a low value is based on the
findings of \citet{CS95}, who argue that when the quasar UV ionizing
background is  the only source of ionization of the HI gas, as should be the
case for dark galaxies with no supernova, then the sound speed of the gas
cannot be  higher than 2 km s$^{-1}$.  Measured values of the sound speed in
the Milky  Way and in nearby galaxies (see e.g., \citet{Kennicutt_89}) range
between 3 and  10 km s$^{-1}$. Note that using a low value for $c_{\rm s}$ is
a conservative assumption, since for small values of the velocity dispersion
the  formation of dark galaxy is more difficult.

More specifically, for a grid in $M_{200}$, $\lambda'$, $f_d$, we  compute the
surface density of the exponential disc $\Sigma$ and whether $Q>1$.   The
shaded area in the left panels of fig.~\ref{fig:darkbestfits} are  obtained
assuming a fixed mass of $5 \times 10^{11}$ M$_{\odot}$, approximately  in the
middle of the mass range considered.  The shaded area in the right  panels is
obtained as follows: we assume a log-normal distribution of spin  parameters
with mean 0.042 and dispersion 0.5 \citep{BDKKKPP01}, thus 68\% of  galaxies
are expected to have $\lambda'>\lambda'_{68}\equiv 0.033$. We then  assume
that a combination of $f_d$ and $M_{200}$ parameters correspond to a dark
galaxy when for all $\lambda'>\lambda'_{68}$ the disc is Toomre stable.  The
Toomre criterion seems to work better for the NFW profile than for the
pseudoisothermal.

However, for the pseudoisothermal profile model there is no theoretical
prediction for the expected $\lambda'$ distribution and the dependence   of
the concentration parameter $c\equiv R_{200}/r_c$ on the halo mass. To
produce the gray areas in fig.~\ref{fig:darkbestfits} we assume the same
log-normal distribution for $\lambda$ as predicted from CDM simulations. This
assumption should be valid since JVO02 find that the empirically recovered
$\lambda'$ distribution is well approximated by the CDM prediction. For
setting  the concentration parameter we consider that, when a galaxy rotation
curve  is fitted by  the pseudoisothermal profile, $r_c$ obtained is about 1/10
of that  obtained when the curve is fitted by the NFW profile (fig. 5 of
JVO02). Thus,  for a given mass, we compute the concentration parameter  as
expected in the  NFW profile, and convert it to the pseudoisothermal  one.
The scatter  around this relation is, however, quite large, thus  introducing
some uncertainty in  the grey areas so obtained.

The Toomre criterion is a local instability criterion; global instabilities
can in principle trigger star formation. For example discs in which the self
gravity is dominant are likely to be unstable to the formation of a bar. Here
we use the results of \citet*{ELN82} and \citet*{MMW98} to conclude that the
onset of these instabilities does not affect the grey areas.

We conclude that the Toomre criteria at least in the context of a NFW profile
seems to predict correctly the region of the parameter space  that galaxies
tend to avoid (i.e., these galaxies should be dark).

\section{Discussion and Conclusions}

In this paper we propose that gas in a large fraction of low mass dark  matter
halos may form Toomre stable disks. Such halos would be stable to star
formation and therefore remain dark. This may potentially explain the
discrepancy between the predicted and observed number of dwarf  satellites in
the Local Group, as well as the deviation between the predicted and the
observed faint end slope of the luminosity function. We show that model  fits
to rotation curves are consistent with this hypothesis: none of the  observed
galaxies lie in the region of parameter space forbidden by the Toomre
stability criterion. Such Toomre stable disks may be the origin of  damped
Ly$\alpha$ systems seen at high redshift  \citep{kauffmann96,JPMH98,MMW98}.

Most semi-analytic models of galaxy formation achieve a reconciliation
between the observed and predicted abundance of low luminosity galaxies by
drastically decreasing $f_{d}$ for faint galaxies. At present, there is no
evidence  from rotation curve modelling that low circular velocity disks are
dark  matter dominated (which would be the case if $f_{d}$ were very
small). Our  model only requires a very gentle decline in $f_{d}$ toward low
masses. It  magnifies the effect of previously proposed mechanisms
(e.g. supernovae feedback, suppression of accretion), since $Q \propto
f_{d}^{-1}$ and low $f_{d}$  disks are more likely to be Toomre stable; {\em
such mechanisms can therefore  be 'tuned down' to lower levels}. This may be
relevant to the problem of  trying to simultaneously fit the luminosity
function and the Tully-Fisher  relation.  In our model, the mass-to-light
ratio is much more stable as a function  of halo mass (except in dark halos,
where it is infinite). Indeed, by construction we fit both the Tully-Fisher
relation and the luminosity function, since both were used to obtain the
velocity function.

\citet{JHHP97} proposed that disks in high spin halos may not be low  surface
brightness galaxies but instead be completely dark. This was at  variance with
the proposal that a large fraction of the 'missing galaxies' lie in low
surface brightness galaxies which formed preferentially in high spin  halos
\citep*{DSS97}. Here, we have extended the \citet{JHHP97} model to a  LCDM
scenario and show that it is supported from an analysis of rotation  curves.
If it is true that disks in such high spin halos may not be low surface
brightness but instead be completely dark, then there should be a  cutoff in
the surface brightness distribution of galaxies; as surveys probe
increasingly low surface brightness galaxies the luminosity density will not
continue to increase but instead will plateau.

It is important to consider independent indications for the existence of this
dark galaxy population and devise possible observational tests.  For example,
\citet{AW01} and \citet*{TMR01} have investigated the detectability of dark
objects.

These dark galaxies have very low surface densities of HI (below 10$^{20}$
cm$^{-2}$ at one scale length of the disk), still high enough to be detected
by current surveys \citep{Zwaan01} if the HI was in its neutral phase.
Current observations that are sensitive enough to detect such amount of
neutral HI have been carried out recently by
\citet{Charlton+00,ZwaanBriggs00,Zwaan01}.  In particular, \citet{Zwaan01}
found no significant detection of HI, down to a detection limit of $7 \times
10^6$ M$_{\odot}$, not associated with visible galaxies. Therefore, it seems
not likely that large masses of neutral HI are harbored in dark galaxies.
This should not come as a surprise, since, given the low surface densities,
the hydrogen might be ionized by the extragalactic background. At densities
below 10$^{19}$ cm$^{-2}$, HI is not able to self-shield from the external
radiation field and most of it will be in its ionized phase.  Another
possibility is that, due to pressure ram stripping from the hot gas of the
hosting large galaxy, most of the HI gas could be removed from the dark galaxy
\citep*{QMB00}.

A more promising route to detect the presence of dark galaxies is through
their gravitational lensing properties. Recently
\citet{MZ01,DK01,Keeton01,Bradac+01} have discussed the need of substructure
to explain the relative fluxes of multiple lensed systems. They show that the
flux ratios observed in each lens can only be explained by the presence of
substructure within a large smooth halo. This population would be similar to
our dark galaxies.

Dark galaxies with dark halo masses above $10^{10}$ M$_{\odot}$ should  be
directly visible via strong gravitational lensing. They will produce splitting
angles $ \lap$ 0.1 arc second, which is just below detection  with current
technology, but will be easily observable with future radio interferometers
(e.g., VLBI)

Using the metallicity distributions of globular clusters,  \citet*{CWM01}
conclude that the mass spectrum of proto-galactic fragments for the  galaxies
in their sample has a power law with index $\sim -2$, indistinguishable  from
that predicted from N-body simulations. They argue that the missing  satellite
population must therefore belong to a class of dark galaxies, similar  to the
ones we consider here.

The biggest caveat in our model is the assumption of conservation of  angular
momentum. In principle angular momentum loss is possible due to  torquing by
substructure in the dark matter halo; indeed, it is invariably seen in  SPH
simulations of disk galaxy formation. This results in more compact  disks
which are much more susceptible to gravitational instability.  However, both
fits to the scale lengths of disks \citep*{MMW98} and modeling of the rotation
curves \citep*{JVO02,vanBBS01} recovers spin parameters consistent with the
log-normal distribution seen in dissipationless N-body simulations,
indicating little or no loss of angular momentum. In addition, we have ignored
the  effect of mergers; as halos merge their disks are likely to be disrupted
and transformed into spheroidal systems, which again are more likely to
become self-gravitating. The merger history of halos is well handled in
semi-analytic models of galaxy formation which employ Press-Schechter based
Monte-Carlo merger trees. It would be very interesting to incorporate the
additional physics of Toomre stability into pre-existing semi-analytic (or
hydrodynamic) galaxy formation models to compare its importance against
various other proposed schemes for suppressing star formation in low mass
halos.

\section*{acknowledgments}
We thank David Spergel for insightful comments and stimulating discussions.
LV is supported in part by NASA grant NAG5-7154. SPO is supported by NSF grant
AST-0096023.  LV and RJ thank the TAPIR group at Caltech for hospitality.

\bibliographystyle{mn2e.bst} 
\bibliography{../../STY/raul}

\end{document}